\documentclass[sigconf,nonacm]{acmart}
\def\BibTeX{{\rm B\kern-.05em{\sc i\kern-.025em b}\kern-.08em
    T\kern-.1667em\lower.7ex\hbox{E}\kern-.125emX}}

\usepackage{enumerate}
\usepackage[shortlabels, inline]{enumitem}
\usepackage{multirow}
\usepackage{multicol}
\usepackage{balance}
\usepackage{booktabs}

\usepackage{listings}
\usepackage{pifont}
\newcommand{\cmark}{\ding{51}} 
\newcommand{\xmark}{\ding{55}}

\usepackage[english, status=final, margin=false, inline=true]{fixme}
\usepackage{appendix}
\usepackage{fancyvrb}
\usepackage{xcolor}
\usepackage{tablefootnote}
\fxusetheme{color}
\FXRegisterAuthor{jf}{ajf}{\color{blue}JF}

\fxusetheme{color}
\FXRegisterAuthor{fn}{afn}{\color{red}FN}

\fxusetheme{color}
\FXRegisterAuthor{yz}{ayz}{\color{violet}YZ}

\fxusetheme{color}
\FXRegisterAuthor{ks}{aks}{\color{orange}KS}

\newcommand{\added}[1]{{\color{black}#1}}

\newcommand{\addedd}[1]{{\color{black}#1}}

\newcommand{\rmspace}{\vspace{-0.0ex}}

\definecolor{goldenrod}{RGB}{245,186,55}
\definecolor{marinblue}{RGB}{57,104,189}
\definecolor{redorange}{RGB}{235,114,40}

\usepackage{balance}

\begin{document}
\title{Evaluating the Data Model Robustness of Text-to-SQL Systems Based on Real User Queries}

\author{Jonathan Fürst}
\affiliation{%
  \institution{Zurich University of Applied Sciences}
  \streetaddress{}
  \city{}
  \country{Switzerland}
  \postcode{}
}
\email{jonathan.fuerst@zhaw.ch}

\author{Catherine Kosten}
\affiliation{%
  \institution{Zurich University of Applied Sciences}
  \streetaddress{}
  \city{}
  \country{Switzerland}
  \postcode{}
}
\email{catherine.kosten@zhaw.ch}

\author{Farhad Nooralahzadeh}
\affiliation{%
  \institution{Zurich University of Applied Sciences}
  \streetaddress{}
  \city{}
  \country{Switzerland}
  \postcode{}
}
\email{farhad.nooralahzadeh@zhaw.ch}

\author{Yi Zhang}
\affiliation{%
  \institution{Zurich University of Applied Sciences}
  \streetaddress{}
  \city{}
  \country{Switzerland}
  \postcode{}
}
\email{yi.zhang@zhaw.ch}

\author{Kurt Stockinger}
\affiliation{%
  \institution{Zurich University of Applied Sciences}
  \streetaddress{}
  \city{}
  \country{Switzerland}
  \postcode{}
}
\email{kurt.stockinger@zhaw.ch}

\begin{abstract}
Text-to-SQL systems (also known as NL-to-SQL systems) have become a popular solution for bridging the gap between user capabilities and SQL-based data access. These systems translate user requests in natural language to valid SQL statements for a specific database. Text-to-SQL systems have benefited from the rapid improvement of transformer-based language models.
However, while Text-to-SQL systems that incorporate such models continuously reach new high scores on---often synthetic---benchmark datasets, a systematic exploration of their robustness towards different data models in a real-world, realistic scenario is missing.

This paper provides the \emph{first in-depth evaluation of the data model robustness of Text-to-SQL systems in practice} based on a multi-year international project on Text-to-SQL interfaces. \added{
In contrast to prior works evaluating Text-to-SQL systems on often synthetically generated generalized multi-domain benchmarks, we evaluate different data models in one domain based on \emph{real-user questions.}

} 
Our evaluation is based on \emph{FootballDB}, a system 
that was deployed over a 9-month period in the context of the FIFA World Cup 2022, during which about 6K natural language questions were asked and executed.
We manually labeled a \added{diverse} subset of these questions for three different data models. \emph{For each data model}, we explore the performance of representative Text-to-SQL systems and language models. We further quantify the impact of training data size, pre- and post-processing steps as well as inference time. Our comprehensive evaluation \emph{sheds light on the design choices of real-world Text-to-SQL systems and their impact on moving from research prototypes to real deployments}.
Last, we provide a new benchmark dataset, which is the first to \emph{enable the evaluation of different data models for the same dataset} and is substantially more challenging than most previous datasets in terms of query complexity.
\end{abstract}

\maketitle

\section{Introduction}
Since the inception of relational databases, SQL has served as the primary query language for structured data sources, offering a declarative interface for application developers to access information. 
In the last decades, SQL has gained importance beyond application developers, as
businesses move to data-driven decision making in the context of data warehouses and data lakes: Cities tackle complex challenges such as carbon reduction~\cite{furst2021realizing}, and scientists in various domains, from social to natural sciences are increasingly following data-driven methods in their research.
This development has led to a shift, in which non-programmers represent a large group of (potential) SQL users. For these users, the SQL interface and the required knowledge of the relational model constitute an often impervious barrier~\cite{ahadi2015quantitative}.

To address the gap between user capabilities and data access, a growing number of Text-to-SQL (also known as NL-to-SQL) systems have been developed in the last years~\cite{affolter2019comparative, gkini2021depth, lehmann2022building, gu2023few}. The idea is compelling: Instead of writing SQL queries, users write queries in natural language and a Text-to-SQL system, with access to the underlying database, translates them into valid SQL statements.
Over the years, there have been various methods from early, keyword and rule-based systems~\cite{blunschi2012soda} to the more recent deep learning-based approaches~\cite{brunner2021valuenet, scholak2021picard} that have pushed SQL generation accuracy on common benchmarks such as Spider~\cite{yu2018spider} to 91.2\% (as of November 2023~\cite{gao2023text}). 
At the core of this success, lie the advancements in transformer-based language models, from Bert~\cite{devlin2018bert} (340M parameters) 
and Bart~\cite{lewis2020bart} (148M parameters),
to T5~\cite{raffel2020exploring} (3B parameters) to the advent of Large Language Models (LLMs), such as OpenAI's ChatGPT~\cite{openai-chatgpt} and GPT-4~\cite{openai2023gpt4}, Google's Bard~\cite{google-bard} or Meta's LLaMA2~\cite{touvron2023llama2} (up to 100s of billions of parameters).

While Text-to-SQL systems incorporating these models continuously reach new high scores on---often synthetic---benchmark datasets, enterprise-grade Text-to-SQL is still far from being resolved as recently noted in~\cite{floratou2024nl2sql}.
\emph{In fact, a systematic exploration of their design space through a real-world deployment with actual user queries does not exist.}
The available survey and analysis papers on Text-to-SQL systems~\cite{affolter2019comparative, katsogiannis2023survey} mainly provide a taxonomy, the needed steps and advantages as well as disadvantages of various approaches.
However, they do not experimentally explore the design space, specifically the data model robustness of any deployed Text-to-SQL systems with real users.
Likewise, many instrumental benchmarks have been proposed for evaluating Text-to-SQL over the years (e.g., Spider~\cite{yu2018spider}, KaggleDBQA~\cite{lee-etal-2021-kaggledbqa}, BIRD~\cite{li2023can}), but they are not constructed from real-world user queries and do not provide the possibility to compare the impact of  different data model design choices on the same set of queries.

In this paper, we provide the \emph{first in-depth design evaluation of the data model robustness of Text-to-SQL systems} based on a multi-year international project on natural language interfaces for databases~\cite{amer2022inode}. \added{Because previous benchmarks and research have been solely fixated on the generalization capabilities of Text-to-SQL systems, in this work, we focus on a single domain \emph{Football} to interrogate other critical aspects of the Text-to-SQL design space.}
Through this exploration, we:
\begin{enumerate*}[label=(\roman*), itemjoin={{, }}, itemjoin*={{, and }}]

\item \emph{Investigate the effect of a growing amount of labeled training data on the accuracy of translating user questions to SQL}
\item \emph{Examine the impact of data model design choices on the results of translating user questions to SQL}
\item \emph{Analyze the role of query complexity in the process of translating user questions to SQL}
  \item \emph{Evaluate Text-to-SQL inference time for different systems}
\end{enumerate*}

We experimentally evaluate these questions using FootballDB---a newly created open dataset with FIFA World Cup data. FootballDB contains \emph{information about world cup games over a 100-year time period}. We collected and curated the dataset semi-automatically from several open data sources (e.g., DBpedia~\cite{auer2007dbpedia}, Wikidata~\cite{vrandevcic2014wikidata}) and through web scraping (e.g., FIFA website). We modeled the data in three different data models: starting from a widely used football data model and two adaptations based on our deployment experience and user feedback. 

Unlike other public Text-to-SQL datasets, \emph{our labeled and evaluated queries have been asked by real users of a deployed Text-to-SQL system several months before and during the FIFA World Cup 2022}. Our system has been queried by several hundred users that issued about 6K natural language questions, 400 of which we have manually labeled for three data models (1200 NL/SQL pairs). \added{The FootballDB dataset adds a fresh outlook to the landscape of Text-to-SQL benchmarks, providing new insights to practitioners who want to deploy these systems on their own databases and researchers who can now evaluate the data model robustness of their developed systems.}

Overall, this paper makes the following \underline{contributions}:

\begin{itemize}
    \item It offers a \emph{comprehensive analysis and definition of the design space for Text-to-SQL solutions}, namely (1) Data Model, (2) Language Model, (3) Training Data Size and (4) Pre-/Post-processing (Section~\ref{sec:design-space}). 
    \item We provide unique insights into \added{\emph{the challenges and requirements of collecting real-user data as well as data model design decisions}} (Sections ~\ref{sec:experience-deployment-Text-to-SQL-system} and~\ref{sec:NL_driven_data_model}) for Text-to-SQL practitioners.
    \item We conduct a \emph{rigorous experimental evaluation of all design space dimensions}, employing state-of-the-art Text-to-SQL systems using small, medium, and large language models along with a realistic, novel dataset and user queries derived from several months of live deployment (Section~\ref{sec:evaluation}).
    \item Drawing from our deployment experience and experimental results, we suggest promising avenues for \emph{future research, focusing on the significance of data models and training data composition}, opposed to merely expanding the volume of labeled data (Section~\ref{sec:discussion}).
    \item Last, we \emph{release FootballDB, a new realistic benchmark dataset} with three data model
    instances with end-user generated queries to the community:\ \url{https://github.com/jf87/FootballDB}
\end{itemize}
\section{Design Space of Text-to-SQL Systems}
\label{sec:design-space}

\textbf{Task Definition.}
Text-to-SQL systems translate the provided natural language question into a
corresponding SQL query. We formally define the process as
follows~\cite{qi2022rasat, li2023graphix, wang2020rat}: Given a natural language question
as a sequence of natural language tokens \added{$\mathcal{Q}:q_1  q_2  \ldots  q_{|\mathcal{Q}|}$} and a relational database schema with
\added{$\mathcal{S}=\{\langle\mathcal{T}_1, \mathcal{C}_1\rangle, \dots, \langle\mathcal{T}_n, \mathcal{C}_n\rangle\}$ where
$\mathcal{T}_i$ is a table,
and 
$ \mathcal{C}_i$ denotes it's corresponding columns (i.e. DB content)},
a Text-to-SQL system is a function $f(\mathcal{Q}, \mathcal{S})$ that outputs
the correct SQL query \added{$\mathcal{Y}:y_1 y_2 \ldots y_{|\mathcal{Y}|}$} as a sequence of tokens.

\subsection{Deep Text-to-SQL Systems}

\begin{figure*}
    \centering
    \includegraphics[width=1.0\textwidth]{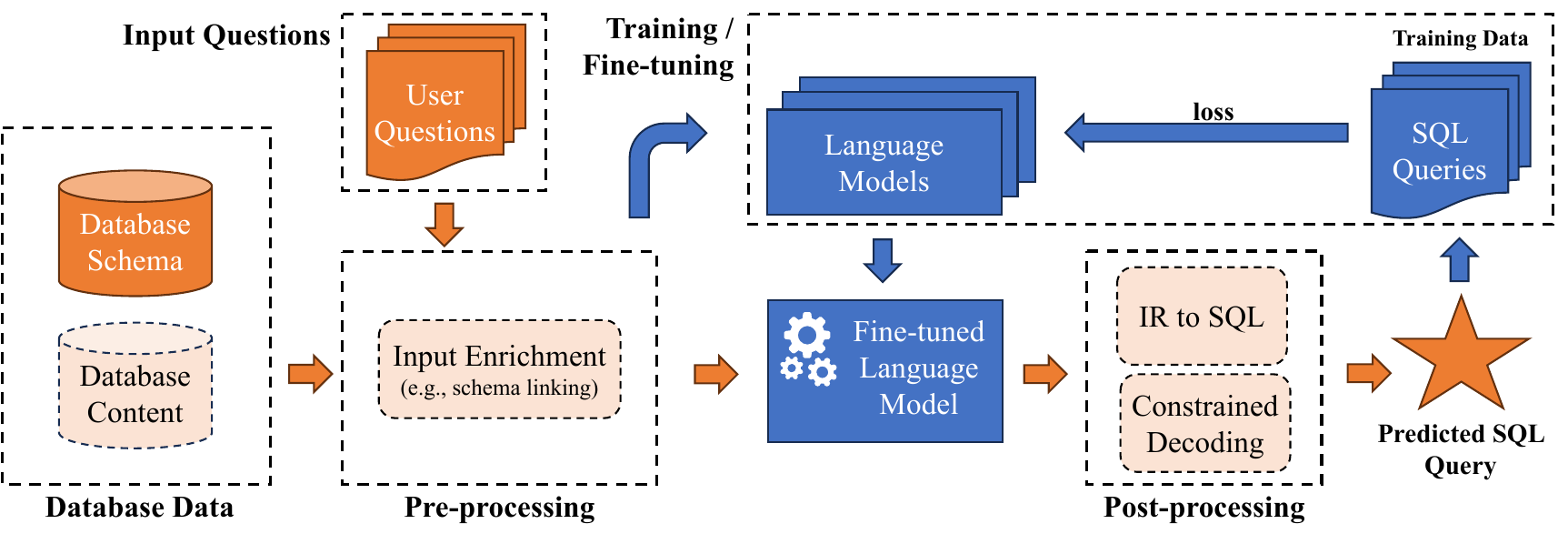}
    \caption{Design space of deep learning-based Text-to-SQL methods (fine-tuning phase in \textcolor{blue}{blue}, prediction phase in \textcolor{orange}{orange}). NL questions, database schema and potentially content serve as input. Pre-processing: Some methods perform input enrichment before the encoded input is fed into a fine-tuned language model. Post-processing: Some methods apply techniques such as intermediate representation (IR) or constrained decoding to improve the quality of the generated SQL query. }
    \label{fig:design_space}
\end{figure*}

State of the art Text-to-SQL systems generate the SQL query $\mathcal{Y}$ with the help of Language Models (LMs)~\cite{brunner2021valuenet, yu2018typesql, guo2019towards}. LMs are transformer-based~\cite{vaswani2017attention}, pre-trained sequence-to-sequence models. Generally, these models process $\mathcal{Q}$, the sequence of natural language tokens through layers of self-attention and feed-forward networks to capture contextual information, and then generate SQL output tokens $\mathcal{Y}$. While, LMs constitute the core functionality of deep Text-to-SQL systems, published methods have added model fine-tuning and various pre- and post-processing steps to customize them for the SQL generation task (see Figure~\ref{fig:design_space}).

\textbf{Pre-processing.} 
In pre-processing, NL questions need to be encoded as tokens $\mathcal{Q}$ together with database schema information $\mathcal{S}$. In addition, some methods also use the database content (e.g., ValueNet~\cite{brunner2021valuenet}) and/or perform input enrichment (e.g., schema linking in IRNet~\cite{guo2019towards} and RAT-SQL~\cite{wang2020rat}). Schema linking connects NL question entities to database tables, columns, and cell values.

\textbf{Training/Fine-tuning.} LMs have been pre-trained extensively on large text corpora (usually obtained from Internet sources). As such, they are not optimized for Text-to-SQL translations.
To provide Text-to-SQL specialization, methods often apply two steps: (1) The model is pre-trained with a large corpus of NL/SQL-pairs (e.g., obtained from a public dataset such as Spider~\cite{yu2018spider}); (2) The model is then further fine-tuned with a smaller set of database-specific NL/SQL-pairs.

\textbf{Post-processing.} Last, the predicted tokens  $\mathcal{Y}$ are converted to a sequence of words, representing a SQL query. Some systems (e.g., IRNet and ValueNet) further use an intermediate representation (IR) such as SemQL which they then convert to SQL. IRs can help bridge the gap between the context free SQL grammar and natural language~\cite{guo2019towards}.
E.g., SemQL eliminates SQL \path{GROUPBY}, \path{HAVING} and \path{FROM} clauses, and conditions in \path{WHERE} and \path{HAVING} are uniformly expressed in the subtree of \path{Filter} in the SemQL query~\cite{guo2019towards}.
While reducing the gap between NL and SQL, the conversion from SemQL to SQL is not lossless. E.g., when a query involves multiple tables, join conditions are chosen heuristically.
NatSQL~\cite{gan-etal-2021-natural-sql} is another widely used IR with a wider range of supported SQL queries. 

A different approach to deal with the gap between NL and SQL has been proposed by Picard~\cite{scholak2021picard}.
As LMs have an unconstrained output space, they can produce invalid SQL statements. Picard thus introduced an incremental parsing method that constrains the decoding outputs in LMs to valid SQL.

\subsection{Design Space Dimensions}
\label{sec:design_space_dimensions}

Based on our observations, we derive the following important design space dimensions for Text-to-SQL systems: 
\begin{enumerate}[start=1,label={[\bfseries D\arabic*]}, leftmargin = 3em]

\item \emph{Data Model.} The database data model (database schema) is one of the main inputs for many Text-to-SQL systems. Furthermore, the design of the data model directly influences the complexity of the corresponding SQL queries. On the one hand, for the relational model, the database community has---through many years of theoretical and practical research---defined design practices (e.g., entity-relationship model~\cite{chen1976entity}), normalization theories and techniques that reduce data redundancy and improve data integrity~\cite{codd1972further}.
On the other hand, for Online Analytical Processing (OLAP) in data warehouses, the star schema~\cite{chaudhuri1997overview} is a data model optimized for efficient data analysis based on de-normalization. \emph{Currently the effect of data models on Text-to-SQL performance is completely under-explored.}

\item \emph{Language Model.} A wide set of increasingly larger trans-former-based language models have been proposed in recent years~\cite{devlin2018bert, affolter2019comparative, openai-chatgpt, openai2023gpt4, google-bard}. These pre-trained language models not only differ in size (from millions to billions of parameters), but also in terms of their availability (open-source, on premise vs. closed-source, cloud-only) and required training/fine-tuning and run-time costs. \emph{The trade-offs of different language model sizes in terms of prediction performance and required costs have not been explored for Text-to-SQL thus far.}

\item \emph{Training Data Size.} The size of the training data used for fine-tuning clearly impacts the performance of the final machine learning model. Surprisingly, the training data size dimension has not been systematically investigated for Text-to-SQL systems. While some have explored domain generalization~\cite{li2023graphix}, to our knowledge, all existing methods have been evaluated on fixed-sized train and test datasets (e.g., from popular benchmarks such as Spider~\cite{yu2018spider} and WikiSQL~\cite{zhongSeq2SQL2017}).
However, when using \emph{Text-to-SQL systems in practice, there is an important trade-off between the effort and cost spent creating labeled training data and the performance.}

\item \emph{Pre- and Post-processing.} Various contributions have been made in the pre- and post-processing steps of Text-to-SQL systems (e.g., input enrichment and intermediate representation~\cite{guo2019towards, brunner2021valuenet} or constrained decoding~\cite{scholak2021picard}). \emph{Currently it is not clear how pre- and post-processing influences the performance of Text-to-SQL translation for different data models.}

\end{enumerate}
\section{FootballDB Overview}

FootballDB serves as a \emph{novel, challenging benchmark for evaluating Text-to-SQL systems based on real user questions and with regards to different data models.}
First, we describe how we construct FootballDB, a dataset comprising almost 100 years of football world cup data. Second, we describe the deployment of FootballDB live before and during the FIFA World Cup 2022. 

\begin{figure}[ht]
    \centering
    \includegraphics[width=1.0\linewidth]{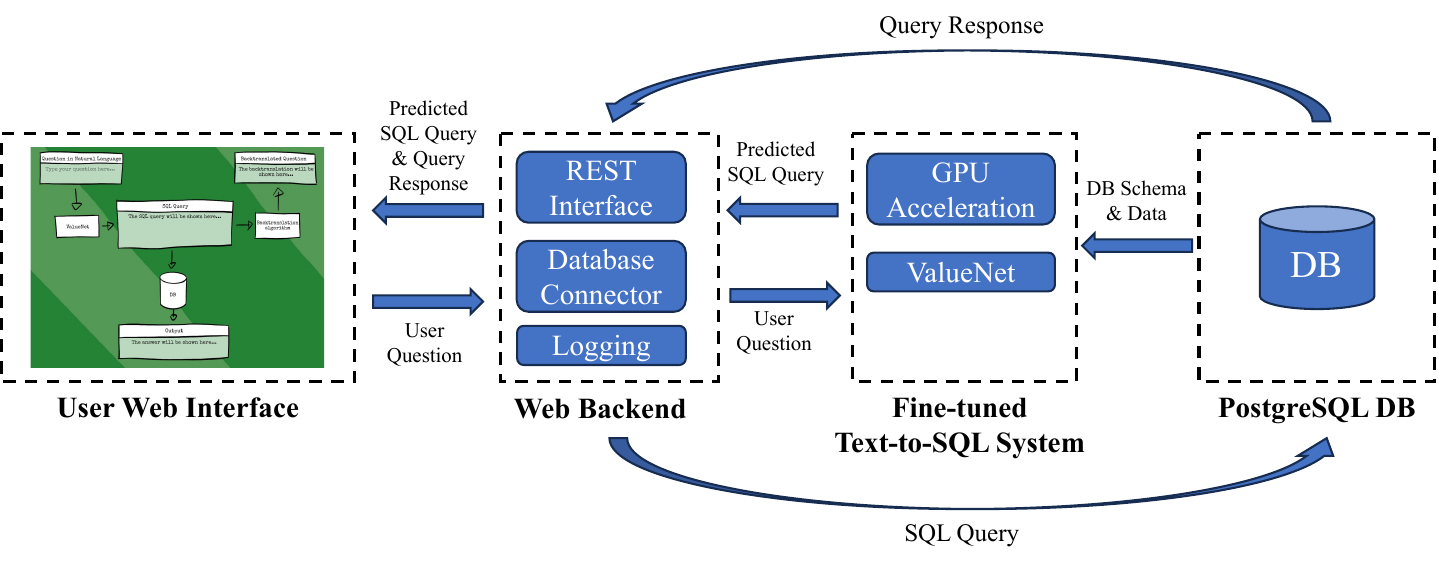}
    \caption{Implementation of the Text-to-SQL system deployed before, during and after the FIFA World Cup 2022.}
\label{fig:worldcup-implementation}
\rmspace
\end{figure}

\subsection{FootballDB Dataset}
The FootballDB dataset is an extensive world cup dataset with information about football games, players, national teams, and clubs dating from the first world cup in 1930 in Uruguay, to the most recent 2022 world cup in Qatar. Football world cups typically take place every four years over a time period of about one month. The number of participating teams has risen from 13 teams in the inaugural World Cup to 32 teams in the most recent edition. 
In total, we collected information about 22 world cups, 86 national teams (including former nations, e.g., the Soviet Union), 8,891 players, and 1,874 clubs.

\textbf{Original Data.}
We sourced the initial dataset from the Kaggle FIFA World Cup dataset~\cite{kaggle-worldcup}. This initial dataset includes 3 CSV-files describing the elimination round matches, players, and final playoffs. However, it only covered the years from 1930 to 2018, contained several erroneous data points, and lacked information about players and national clubs.

\textbf{Data Enrichment and Cleaning.}
To provide a comprehensive view of the World Cup and address frequent user queries about the clubs of players, we supplemented the dataset with additional information on leagues, clubs, and coaches from Wikidata.
We also performed extensive manual checks to ensure high-quality and accurate data. For example, we included the full names of players (previously, only nicknames of partial names were used) and their dates of birth to de-duplicate and augment player data. We also correct inaccuracies and added missing players for each participating country's team. In total, we added 1,230 new players, 89 leagues, 1,874 clubs, and 1,966 coaches to the initial dataset. 

To provide an accurate and updated data source for viewers and fans of the FIFA World Cup, we continuously updated our database after each set of matches throughout the event \added{in 2022}. Specifically, we added new data points for each match, including detailed information about goals, penalties, red and yellow cards, as well as the minutes during the game when each of these events occurred. \emph{FootballDB can be considered the most accurate, comprehensive, open-source relational database for the world cup championship to date}. 

\subsection{FootballDB Deployment}
\label{sec:deployment}

For our deployment, we created a modular architecture for the user web interface, including a web back-end and Text-to-SQL system. Data was stored in a PostgreSQL database (see Figure~\ref{fig:worldcup-implementation}).
During our live deployment, we employed ValueNet~\cite{brunner2021valuenet} as the Text-to-SQL system. We chose ValueNet since the source code is available and it has been used extensively in various projects between academia and industry.

\textbf{Text-to-SQL System: ValueNet.}
We based our deployment on ValueNet, which is built on some of the components of IRNet~\cite{guo2019towards}. The main novelty of ValueNet is a pre-processing method that not only employs IRNet's schema linking mechanism but also extracts values from natural language questions and infers possible value candidates from the base data of a given database, even when not explicitly stated in the natural language question.
ValueNet uses a BART encoder~\cite{lewis2020bart} (148M parameters) and incorporates the database schema and content as input to the neural network.
For post-processing, ValueNet uses an IR (SemQL) for translating natural language to SQL. This approach is limited to the SQL language features supported by the grammar (see also Section~\ref{sec:design_space_dimensions}).

\begin{figure*}[ht]
    \centering
    \includegraphics[width=1.0\textwidth]{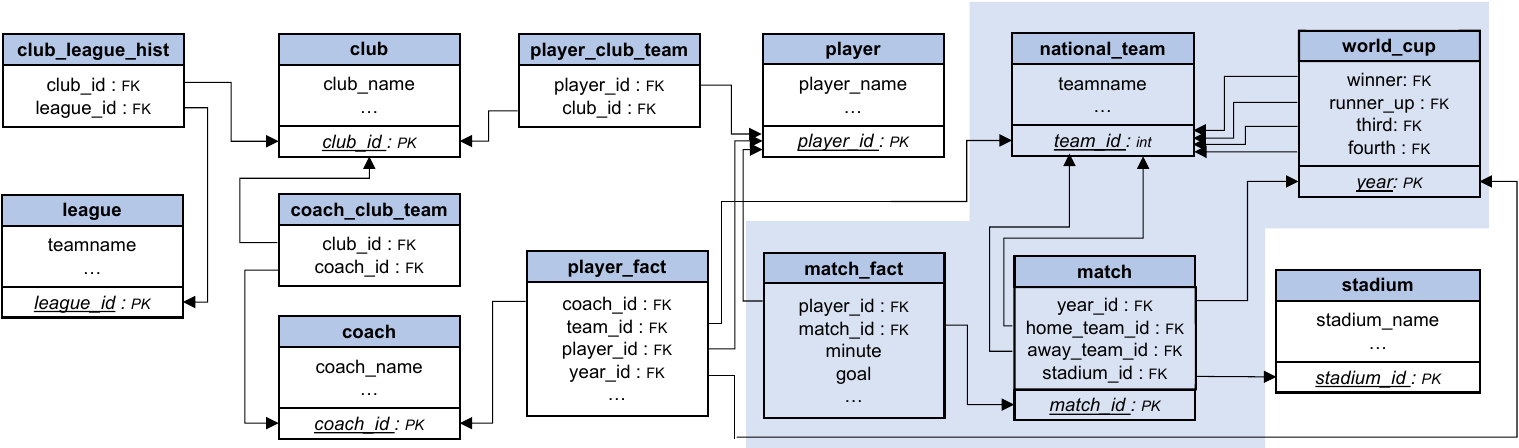}
    \caption{World Cup Schema Diagram v1. PK = Primary Key. FK = Foreign Key. The tables highlighted in blue are affected by re-designing the data model in various iterations. Note that the 1:n relationships between tables \texttt{national\_team} and \texttt{match} as well as between \texttt{national\_team} and \texttt{world\_cup} contain multiple PK/FK references.}
    \label{fig:db_model_v1}
\end{figure*}

\textbf{Initial Data Model.}
We created an initial data model for our FootballDB dataset based on our domain knowledge and commonly applied database modeling practices~\cite{buf03, silberschatz2005database}.
This initial data model contains information about players, matches, stadiums, national teams, clubs, coaches, etc. (see Figure~\ref{fig:db_model_v1}).
Overall, the resulting database schema contains 13 tables with 14 foreign key constraints 

In the following two Sections, we \added{detail the requirements, challenges, and our process of collecting real-user questions
(Section~\ref{sec:experience-deployment-Text-to-SQL-system})} 
and then provide an in-depth discussion of the design decisions that went into the development of our three data models (Section~\ref{sec:NL_driven_data_model}).

\added{\section{Real-User Data Challenges}
\label{sec:experience-deployment-Text-to-SQL-system}
}
Despite the intensive research focused on improving Text-to-SQL performance on existing benchmarks, very little work on \added{how real users interact with systems in the real world exists. }
This section aims to diminish this gap, \added{by reporting the requirements, challenges, and process of collecting NL/SQL pairs through real-user interactions.}

\noindent \added{\textbf{Challenge 1: Attracting Users.}}
A major roadblock in evaluating natural language interfaces for relational databases in the real world is to attract \added{a diverse set of} users. 
To attract a larger user pool, we chose the FIFA World Cup 2022 because of its international attention and diverse audience. Football is a topic that many people, regardless of their background, education, or culture, can ask questions about. 
\added{We advertised the deployment within the university and with our industry partners and promoted it during citizen-research engagement events and to first-year computer science undergraduates.}
From the beginning of the live deployment in October 2022 until the end of the world cup in December 2022, we collected around 6K queries from several hundred users. 

\noindent \added{\textbf{Challenge 2: Identifying and Fulfilling User Needs.}}
\added{A second challenge is to create a system that fulfills the information needs of users by providing answers to their football-related questions.
Therefore, we continuously analyzed the posed questions to discover emerging topics that could not be answered by the content of the database. 
}
The questions showed a trend in user interest in the clubs that players had played in, which coaches had coached the players, and which leagues were currently division one. Based on this feedback \emph{we added new data to our database} about clubs, leagues, players, and coaches. 

\noindent \added{\textbf{Challenge 3: Harnessing User Knowledge.}}
\added{A third requirement is to benefit from the SQL and domain knowledge of users to receive feedback on the quality of the generated outputs. This is important for understanding if the information needs for a specific question are actually fulfilled by the provided answer.} We designed a \emph{new interface for expert users} that are either familiar enough with SQL to recognize that a SQL query is incorrect or expert enough in world cup statistics to know that the output of the query is incorrect.
Users could add a thumbs up or thumbs down, indicating whether a given result is correct or not. SQL experts could edit and correct the SQL query, which was then logged by our system. These additional data points were then used to filter and clean the logged natural language questions and SQL queries.

\noindent \added{\textbf{Challenge 4: Labeling User Data.}}
\added{Last, data labeling is a big bottleneck in most ML systems, usually requiring manual effort~\cite{furst2020towards, furst2023versamatch, cheng2024interactive}. Here, this effort translated to manually verifying and labeling all posed user questions and generated SQL outputs. Therefore, processes to reduce this manual effort are needed.
For queries that had been marked correct by users (see Challenge 3), we manually verified them. For queries that had been marked either incorrect or not at all by users, we developed several automation techniques to speed up the otherwise enormous manual effort:
}
First, we computed the cosine similarity of the embedding vectors of all natural language questions using \textit{SentenceBERT}~\cite{reimers2019sentence} to \emph{compute the semantic textual similarities}. We then set a high similarity threshold ($\geq 96\%$) to automatically label queries that were very similar to previously manually verified queries. 
Second, we also \emph{used this similarity measure as a labeling aid} to show labelers similar to previously validated natural language questions and SQL query pairs ($<96\%$). Having a similar query helped labelers to more quickly and accurately identify errors in queries and correct them, e.g., missing filters, incorrect table joins, missing columns in projections, or missing aggregations. 

\noindent \added{\textbf{Overall Observations.}}
\added{We observed interesting patterns in the posed user questions that should be considered in the design and evaluation of future Text-to-SQL systems: 
1) \emph{unrelated questions}, 2) \emph{unanswerable questions}, (3) ambiguous questions (4) \emph{questions in languages other than English}, and (5) a multitude of \emph{spelling errors} for player names.
These observations are similar to the recently presented open issues in achieving enterprise-grade Text-to-SQL~\cite{floratou2024nl2sql}.
Specifically, their two main problems (1) natural language ambiguity and (2) semantic mismatch (i.e., the user’s intent cannot be effectively
fulfilled by the provided database), can also be directly observed in our collected data.
}
We collected around 6K user interactions (see Table \ref{fig:log_statistics}). For these, our system was able to generate SQL queries 89\% of the time. Failures to generate SQL queries are generally due to: an NL question in a different language, out of scope, lack of training data representing similar questions. 

\begin{table}[ht]
\centering
\small
\begin{tabular}{l || r  }
\toprule
    Type of User Log & Amount of Logs  \\ \hline
    \#NL questions issued & 5,900\\
    \#Times SQL generated & 5,275\\
   \#Times no SQL generated & 625\\ 
    \#Thumbs up & 174\\
    \#Thumbs down & 949\\
    \#\added{User} corrected SQL queries & 1,287\\
\bottomrule
\end{tabular}
     \caption{Statistics of live user logs collected during and after the world cup from November 21, 2022 to December 31, 2022.}
    \label{fig:log_statistics}
    \rmspace
\end{table}
\section{Data Model Design}
\label{sec:NL_driven_data_model}

We discuss two adaptations for improving the initial data model of FootballDB.
Our primary objectives in the design process have been to create a data model driven by two goals:

\begin{enumerate}[start=1,label={[\bfseries G\arabic*]}, leftmargin = 3em]
    \item \emph{User Goal.} We aim to reduce the complexity of SQL queries and support NL questions that are commonly asked by users in our deployment as described in Section~\ref{sec:experience-deployment-Text-to-SQL-system}. 
    \item \emph{System Goal.} We aim to minimize potential errors that originate in the Text-to-SQL system's pipeline, such as in the pre- and post-processing steps or within the language model as described in Section~\ref{sec:design_space_dimensions}.
\end{enumerate}

\subsection{Starting Point: Data Model v1}

Our initial data model (see \textbf{v1} in Figure~\ref{fig:db_model_v1}) consists of 13 tables with 14 foreign key constraints. Note that four tables are highlighted in blue. These are the focus of remodeling as they contain the core information about national teams and matches. This focus is supported by the fact that these tables are accessed in around 74\% of user questions in our deployment.

\begin{figure*}[ht]
    \centering
    \includegraphics[width=1.0\textwidth]{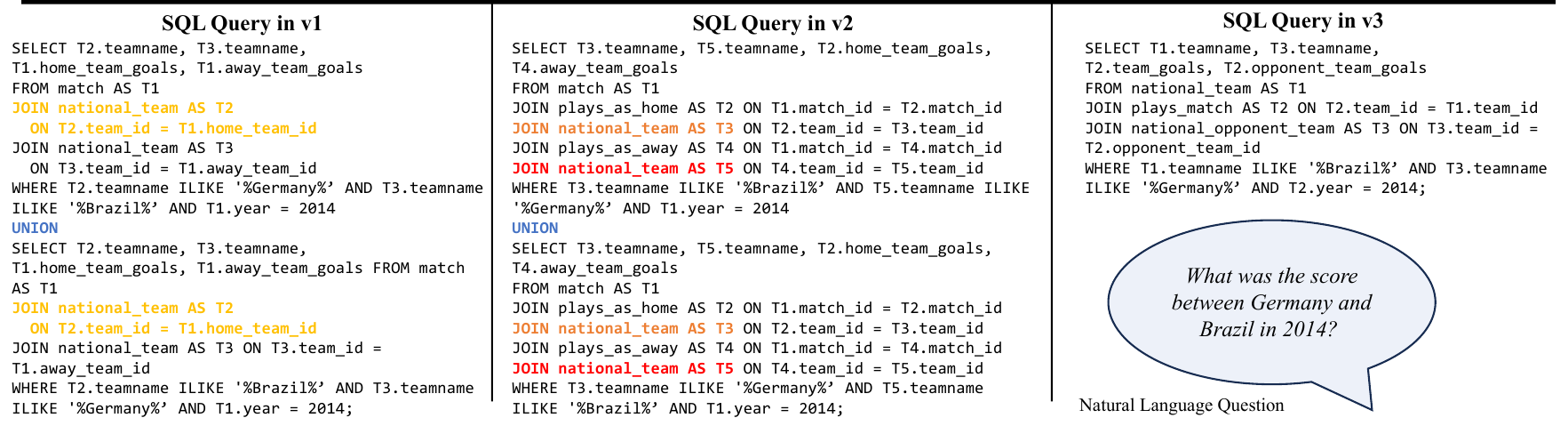}

    \caption{Example SQL query in all three data models of the natural language question: \textit{What was the score between
Germany and Brazil in 2014?} Components shown in \textcolor{goldenrod}{yellow} indicate the unsuccessful interpretation of joins; components shown in \textcolor{marinblue}{blue} indicate set operations; components shown in \textcolor{redorange}{orange} and \textcolor{red}{red} indicate the multiple instantiations of a table in a \texttt{SELECT} clause.}
    \label{fig:sql_sample}
\end{figure*}

As described in Section~\ref{sec:experience-deployment-Text-to-SQL-system}, we started with data model v1 depicted in Figure~\ref{fig:db_model_v1}.
Note the $1:n$ relationship between \path{match} and \path{national_team}. We can see that table \path{match} contains multiple foreign keys that reference the table \path{national_team}. However, this causes problems for Text-to-SQL systems that employ an intermediate representation (IR) such as SemQL as part of their post-processing step. E.g., the shortest path algorithm employed by such systems for generating SQL queries only supports a \emph{single primary key/foreign key references between any two tables} to form the subgraph of the join path \cite{brunner2021valuenet, guo2019towards}.
Hence, the table join-path algorithm fails at the post-processing stage. 
E.g., the SQL query in v1 listed at the left side of Figure~\ref{fig:sql_sample} corresponds to the natural language question \textit{``What was the score between
Germany and Brazil in 2014?''} given data model v1. The \textcolor{goldenrod}{yellow} components of the query represent the joins that cannot be correctly interpreted.
A similar case is a 1:n relationship between \texttt{national\_team} and \texttt{world\_cup} in Figure \ref{fig:db_model_v1}, where \texttt{national\_team} is referenced by four foreign keys. 

During the initial deployment, we found that very few questions asked by users contained keywords related to the terms ``home team'' or ``away team'', defined by \path{home_team_id} and \path{away_team_id} in the table \path{match}. Instead, users preferred to describe matches in terms of one team against another, e.g., Brazil against Germany. As a consequence, the corresponding SQL query is too complex to conform to the intermediate representation format used in the pre-processing step shown in Figure \ref{fig:design_space}. Moreover, the resulting SQL queries may be too long to fit the maximal length of the input tokens (1024) supported by the language model.

\subsection{First Optimization: Data Model v2}
\begin{figure}[ht]
    \centering
    \includegraphics[width=1.0\columnwidth]{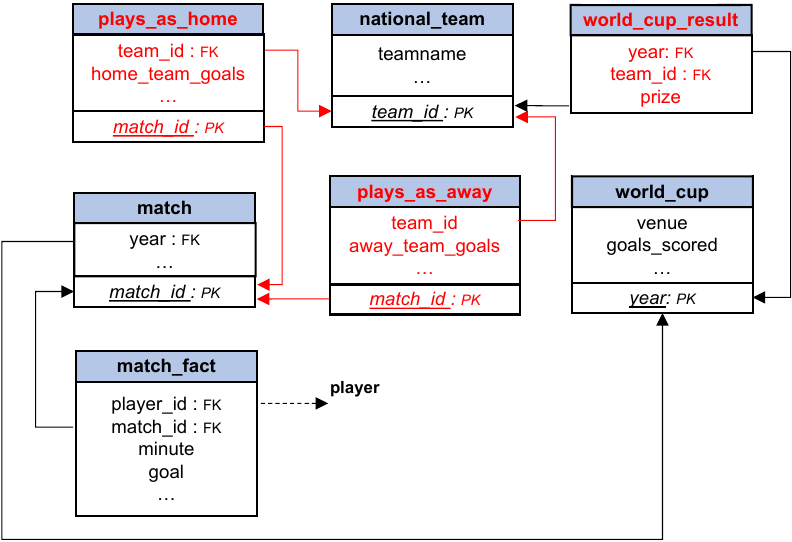}
    \caption{World Cup Schema Diagram Changes v2. Tables highlighted in red are changed with respected to data model version 1 shown in Figure \ref{fig:db_model_v1}. Original 1:n relationships containing more than one PK/FK references are remodeled to only contain a single PK/FK reference between two tables.}
    \label{fig:db_model_v2}
\end{figure}

To tackle the issue of only supporting a single primary key/foreign key reference between two tables, we modified the data model as follows:
The original $1:n$ relationship between \path{national_team} and \path{match} is remodeled with the two bridge tables (also known as \texttt{composite entity types}\footnote[1]{Proposed by the guide of creating correct ER diagram from \cite[p.151-153]{buf03}}) \path{plays_as_home} and \path{plays_ as_away} (see \textbf{v2} in Figure \ref{fig:db_model_v2}). As a result, we eliminate  multiple primary/foreign key references between tables \path{match} and \path{national_team}.
Note that in database schema v1 shown in Figure \ref{fig:db_model_v1}, there are also multiple primary/foreign key references between the tables \path{national_team} and \path{world_cup}. We again solve this issue by remodeling the table \path{world_cup}  with one additional table \path{world_cup_result}.

Following the transition of the database schema from v1 to v2, the system now accurately handles the majority of single join paths.
Nevertheless, a certain amount of queries remain unresolved as they \emph{cannot pass through the Spider SQL parser}~\cite{yu2018spider} in the stage of pre-processing.
This parser is a core component of many Text-to-SQL systems such as IRNet~\cite{guo2019towards}, ValueNet~\cite{brunner2021valuenet} and RAT-SQL~\cite{wang2020rat}.
The parser does not support multiple table instances with different table aliases. These aliases result in an inferred SQL query, of which the join path may contain either \texttt{match\_home\_team} or \texttt{match\_away\_team}, but not both simultaneously. 

This situation requires a workaround through the modification of SQL queries, involving the utilization of the set operation \textcolor{marinblue}{\texttt{UNION}} (see SQL queries in v1 and v2 in Figure~\ref{fig:sql_sample}). However, this leads to a \emph{twofold increase of the SQL query length and adds complexity to the query structure}. Further, when there is no workaround, and the occurrence of multiple instances of a specific table is inevitable, the SQL queries cannot be processed.
In the center of Figure \ref{fig:sql_sample}, the SQL query in v2 also demonstrates this case. The highlighted components (\textcolor{orange}{orange} and \textcolor{red}{red}) indicate the multiple table instantiations that cause the input parser to fail during the pre-processing stage. 

Another challenge is the well-known \emph{lexical problem}~\cite{dong-etal-2019-data}, which is the cause of the poor performance for questions related to the prize columns in table \texttt{world\_cup\_result}. This effect was especially noticeable in the case of the prize of  ``runner-up''. It appears that people prefer to use a more intuitive expression, such as ``second place'' or ``lost in the final''. Through analysis of all user questions, we discovered that these terms are in fact used $\approx 3$ times as often.

\subsection{Second Optimization: Data Model v3}

\begin{figure}
    \centering
    \includegraphics[width=1.0\columnwidth]{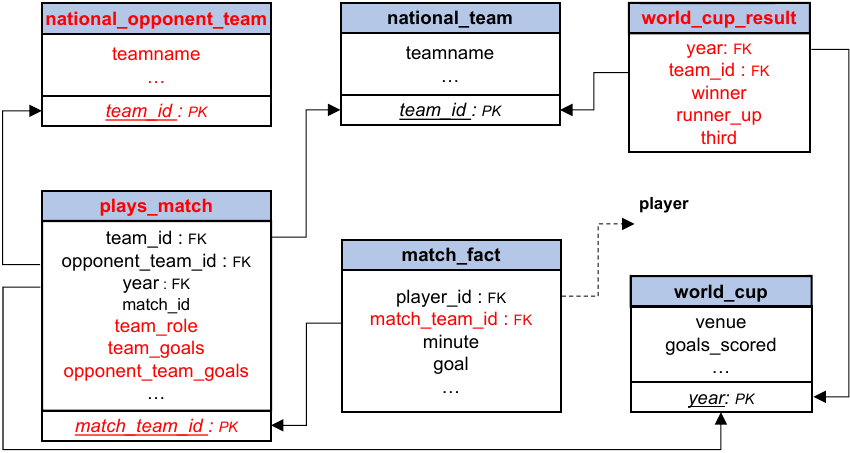}
    \caption{World Cup Schema Diagram Changes v3. Tables highlighted in red are changed with respect to data model version 2 shown in Figure \ref{fig:db_model_v2}.}
    \label{fig:db_model_v3}
    \rmspace
\end{figure}

To address these issues, we introduce the following hypothesis. For humans, query writing becomes easier, if the query
  \begin{enumerate*}
      \item comprises fewer JOINs, which results in a shorter overall query length;
      \item has a more intuitive structure with self-descriptive semantics;
      \item and requires little or no implicit knowledge.
  \end{enumerate*}  
Our hypothesis is that the factors in writing an easy query for humans should also influence deep Text-to-SQL systems, since their trained models aim to mimic human-like query generation capabilities. 
Hence, we modified the data model as follows (see \textbf{v3} in Figure \ref{fig:db_model_v3}). 

\begin{itemize}

\item To simplify writing queries about football matches between different national teams, we removed the non-intuitive tables \path{plays_as_home} as well as \path{national_opponent_team} and \path{plays_match}. Hence, the fact that, for instance, Brazil plays against Argentina can now more intuitively expressed by the query \path{national_team} $\bowtie$ \path{plays_match} $\bowtie$ \path{national_opponent_team} (see also right query in Figure \ref{fig:sql_sample}). 
\item  To implement the proper relationships adapted to the semantics of the natural language question, we redesign the relevant columns and foreign keys\footnote{Removing columns \texttt{home\_team\_id} and \texttt{away\_team\_id}; inserting  \texttt{team\_id} and  \texttt{opponent\_team\_id}}, and add the extra column \texttt{team\_role} to indicate the role of the respective team either as home or as away.
\item Since both teams in a match can be recognized as home or opponent, we adapt all the rows for the new \texttt{plays\_match} table and add a new primary key \path{match_team_id}, which is a concatenation of \path{match_id} and \path{team_id}.
\item To enhance the ability of the language model to create links between the NL questions and the database schema, we expand the table \path{world_cup_result} and convert the single text column, \path{prize}, into four Boolean columns with more intuitive names, \path{winner}, \path{runner_up}, \path{third}, and \path{fourth}.
This optimization simplifies the value extraction phase from searching for entities in both the schema and content of the database, to focus on the schema only. 

\end{itemize}

\textbf{Summary of Data Model Modifications.}
Table~\ref{tab:complexity_overview} shows the major characteristics of the three different versions of our data models and highlights all modified tables and columns.
Note that data model v2 with has the highest number of tables and the lowest mean number of columns per table.
In other words, this data model has a high number of tables, but a low number of columns and low mean number of columns per table. On the other hand, data model v3 has a high number of tables and columns, but it surprisingly simplifies the query writing process (see the observed example query length in Figure~\ref{fig:sql_sample} as well as the mean query length in Table~\ref{tab:train_dev_data_statistics_transposed}).
We will explore the impact of different data models for Text-to-SQL systematically in the next section.

\begin{table}[t]
\centering
\small
\begin{tabular}{l | r r r}
\toprule
            & \textbf{DB v1} & \textbf{DB v2} & \textbf{DB v3} \\ \midrule
\#Tables    & 13 & 16 & 15 \\ 
\#Columns    & 97 & 98 & 107 \\ 
\#Rows    & 104,531 & 106,547 & 106,111 \\ 
\#FKs     & 14 & 13 & 16 \\ 
Mean \#Columns per Table    & 7.46 & 6.13 & 7.13 \\ 
Mean \#Rows per Table    & 8,041 & 6,659 & 7,074 \\ 
\bottomrule
\end{tabular}
\caption{Characteristics of FootballDB across three different data models: Number of tables, columns, rows, etc.}
\label{tab:complexity_overview}
\rmspace
\end{table}
\section{Experimental Evaluation}
\label{sec:evaluation}

We perform a detailed experimental evaluation of different Text-to-SQL systems and data models against our dataset FootballDB.
Our overall goal is to investigate the affect of the identified design dimensions in Section~\ref{sec:design_space_dimensions} through the following research questions (RQs):

\begin{itemize}
    \item RQ 1: \emph{How do different \underline{data models} change the accuracy of Text-to-SQL systems for translating natural language questions to SQL using our novel dataset collected by real users?} \added{$\rightarrow$~Section 6.2}
    \item RQ 2: \emph{What is the accuracy of Text-to-SQL systems that leverage small, medium or large \underline{language models}?} \added{$\rightarrow$~Section 6.2}
    \item RQ 3: \emph{How does \underline{training data size} impact Text-to-SQL performance?} \added{$\rightarrow$~Section 6.2: Train Set Size}
    \item RQ 4: \emph{What are the \underline{systematic errors} of Text-to-SQL systems based on query complexity and query characteristics?} \added{$\rightarrow$~Section 6.3}
    \item RQ 5: \emph{What is the \underline{inference time} of Text-to-SQL systems, and can they be used for interactive data exploration where response times of less than 3 seconds are expected?}
    \added{$\rightarrow$~Section 6.4}
\end{itemize}

\subsection{Experimental Setup}
\textbf{Train and Test Set.}
As described in Section~\ref{sec:experience-deployment-Text-to-SQL-system}, we logged 6K real user interactions with our system. 
\added{
We filter these 6K questions to remove duplicates, questions in languages other than English, or questions that were entirely unrelated to the dataset, i.e., questions not about football. We then label the remaining NL/SQL pairs, i.e., we check if the SQL queries correctly expressed the NL questions, rewriting the SQL queries if not. }

\added{
To create a rich and comprehensive dataset, linguistically and in terms of database coverage, we use a clustering-based approach to select a diverse set of NL questions. Specifically, we apply topic modeling using BertTopic~\cite{grootendorst2022bertopic} to create dense clusters of related NL questions. We then sample and label each cluster as follows. First, we sample the centroid NL question and then select all NL questions in the cluster that are below a threshold of 0.93 \addedd{in terms of BERT similarity to the centroid}. This threshold is chosen so that it results in approximately 1K questions, which we manually label for data model v3.
By excluding samples proximal to the cluster centroid, we can effectively reduce the dataset size while minimizing the loss of diversity within the dataset. 
E.g., this method avoids labeling semantic and lexically similar questions such as ``Who won the world cup in 2014?'' and ``Who won the world cup in 2018?''. 

Second, because manual labeling is a considerable effort (e.g., involving six expert annotators), we then only choose a subset of 400 NL/SQL-pairs to label for data models v1 and v2. These NL/SQL-pairs are chosen through uniform sampling from the 1K NL/SQL-pairs based on their SQL query hardness level proposed in the Spider paper~\cite{yu2018spider}.
The Spider SQL query hardness is rule-based and has 4 levels of difficulty: easy, medium, hard and extra hard. Easy queries have a single projection (this can be an aggregation or not) and no joins. Medium queries have more than 1 projection, more than 1 join and can include filters, order by, having or group by clauses. Hard queries have several projections and multi-table joins and can include set operations, multiple filters and order by, having or group by clauses. Extra hard queries have several projections, multi-table joins and can include set operations, subclauses as well as multiple filters and order by, having or group by clauses.
\emph{Overall, our method ensures a diverse sampling from the NL perspective (through topic modeling) and from the SQL complexity perspective (through Spider hardness).
We release the full dataset, including the 6K raw dataset and 1K gold dataset for v3, which allows for a different sample selection strategy in the future.
}
}

\added{Overall this results} in 1,200 manually corrected NL-SQL pairs \added{for data model v1, v2 and v3 (400 each)}.
\addedd{Per data model, we use 100 records as the test set to evaluate the Text-to-SQL accuracy. The remaining 300 records are the train set. For our experiments, we use 100, 200, and 300 records, respectively, to study the impact of increasing the size of the training data on the translation accuracy. The results are shown in Table~\ref{tab:model_accuracies_traditional} and Table~\ref{tab:model_accuracies_llms}}. 

Note that the \emph{query characteristics vary greatly between the models} \addedd{(see Table~\ref{tab:train_dev_data_statistics_transposed})}. For example, due to the modeling change in v3, set operations are no longer necessary. The v2 data model requires the most number of joins, because it contains the most tables. Surprisingly, the widely used Spider hardness level~\cite{yu2018spider} only shows a small increase from v1 to v2, with the lowest value for v3 (we map the Spider hardness levels to numeric values \added{from 1 for easy to 4 for extra hard.}). This might indicate that the metric should be revised or expanded to better reflect other factors, such as the number of joins or the number of SQL tokens as shown in the recently released BIRD benchmark~\cite{li2023llm}. 
Note that FootBallDB queries have around 100\% / 300\% more joins and 10\% / 100\% more SQL tokens per query than the queries in the recent BIRD benchmark and the heavily used Spider benchmark~\cite{yu2018spider}, respectively.

\textbf{Evaluation Metrics.}
A popular evaluation metric for Text-to-SQL systems is the Semantic Evaluation for Text-to-SQL with Test Suites~\cite{zhong2020semantic}. However, this evaluation script is unable to parse some samples of our train and test sets due to the built-in limitations of its SQL parser~\cite{yu2018spider}. 
Therefore, we apply \emph{exact execution matching} (EX), also known as result matching~\cite{kim2020natural} as the accuracy metrics instead of exact SQL component matching as in the test suite evaluation.
EX denotes the fraction of questions within the evaluation set, where the outcomes of both the predicted and ground-truth queries yield identical results relative to the total number of queries.

\begin{table}[ht]
\centering
\small
\begin{tabular}{l | rrr | rrr}
\toprule
 & \multicolumn{3}{c}{\textbf{Train Set}} & \multicolumn{3}{c}{\textbf{Test Set}} \\
\midrule 
\textbf{Data Model} & v1 & v2 & v3 & v1 & v2 & v3 \\
\midrule 
\#Joins & 1.68 & 2.23 & 1.45 & 1.78 & 2.63 & 1.45 \\
\#Projections & 1.81 & 1.81 & 1.85 & 2.07 & 2.08 & 1.96 \\
\#Filters & 1.95 & 2.20 & 1.73 & 2.15 & 2.39 & 1.85 \\
\#Aggregations & 0.48 & 0.49 & 0.48 & 0.58 & 0.59 & 0.53 \\
\#Set Operations & 0.14 & 0.14 & 0.00 & 0.17 & 0.19 & 0.00 \\
\#Subqueries & 0.05 & 0.05 & 0.01 & 0.03 & 0.03 & 0.03 \\
Mean Hardness & 3.03 & 3.06 & 3.00 & 3.10 & 3.18 & 3.02 \\
Mean Query Length & 217 & 252 & 187 & 232 & 282 & 193 \\
\bottomrule
\end{tabular}
\caption{Query characteristics: Mean values of train and test set used for training and evaluating. Hardness corresponds to the classification defined in~\cite{yu2018spider}. Query length is defined as the number of characters per SQL query.}
\label{tab:train_dev_data_statistics_transposed}
\rmspace
\end{table}

\textbf{Text-to-SQL Systems.}
For our evaluation, we select five Text-to-SQL systems summarized in Table~\ref{tab:lm_dimensions} that cover the design space presented in Section~\ref{sec:design_space_dimensions}. Note that with new Text-to-SQL approaches being proposed on a monthly basis,
we aim to select representative systems with the intention that our insights also apply to recent and prospective specialized approaches that might build on the language models and/or pre- and post-processing methods found in our selection.

\emph{Small Language Models.}
We use ValueNet~\cite{brunner2021valuenet} 
which improves upon IRNet~\cite{guo2019towards}.
ValueNet uses BART~\cite{lewis2020bart} as an encoder, which can be considered a first-generation, small language model (148M parameters). It also incorporates additional pre- and post-processing steps in its pipeline (see also description in Section~\ref{sec:deployment}).

\begin{table*}[t]
\centering
\resizebox{1\textwidth}{!}{
\begin{tabular}{c|p{3.4cm} || c | c c|c | c}

\toprule
\multicolumn{2}{c||}{\textbf{Dimensions}} & \textbf{ValueNet} & \textbf{T5-Picard}
& \textbf{T5-Picard$_{Keys}$}
& \textbf{GPT-3.5} 
 & \textbf{LLaMA2-70B}
\\ 
\midrule
\multirow{4}{*}{Language Model}&Scale (\#Params) & small (148M)  & medium (3B)
& medium (3B) & large (175B)  & large (70B) \\
&DB Schema w/ FK & Yes (with) & Yes (without)
& Yes (with) & Yes (with)   & Yes (with) \\
&DB Content & Yes  & No
& No & No & No \\
&
Output Specification & IR  & SQL
& SQL & SQL  & SQL \\

\midrule
\multirow{3}{*}{Pre-processing} & Query Normalization & SQL-Parser & String Normalization  
& String Normalization  & String Normalization & String Normalization\\

&Value Finder & Yes & No
& No & No   & No  \\
&Conversion to IR & Yes  & No
& No & No  & No \\
\midrule
\multicolumn{2}{c||}{Post-processing}& IR to SQL  & Picard
& Picard & N/A  & N/A \\
\bottomrule
\end{tabular}
}
\caption{Characteristics of the Text-to-SQL systems in our experimental evaluation as well as the dimensions of the language models, pre-processing, and post-processing in different Text-to-SQL systems. FK = foreign key; IR = intermediate representation; 
String Normalization = set of string formatting operators which aim to remove tabs, line breaks, multiple consecutive spaces.}
\label{tab:lm_dimensions}
\rmspace
\end{table*}

\emph{Medium Language Models.}
We employ T5-Picard~\cite{scholak2021picard} as 
a representative second-generation model. T5-Picard comprises sequence-to-sequence models~\cite{raffel2020exploring} incorporating encoder and decoder components that are entirely based on transformer architectures, in contrast to the previous encoder-only language models. Picard is a method for constraining auto-regressive decoders of language models through incremental parsing to only valid SQL output sequences. Besides its original implementation, the method is used in four more Text-to-SQL systems currently found on the Spider leader board~\cite{zhao2022importance, zeng2023n, qi2022rasat, li2023graphix}.
Based on our results (see Section 6.2), we also develop one variation leveraging the base T5 model~\cite{raffel2020exploring} and Picard:
T5-Picard$_{Keys}$, which includes primary and foreign key constraints, not included in T5-Picard.

\emph{Large Language Models (LLMs).}
We conduct LLM experiments using OpenAI’s gpt-3.5-turbo~\cite{openai-chatgpt} and LLaMA2-70B~\cite{touvron2023llama2}, an open-source model from Meta. Currently, OpenAIs GPT models are used in the best six systems on the Spider leaderboard~\cite{dong2023c3, gao2023text, pourreza2023din}, while LLaMA2 is considered one of most widely used open-source models.
We set
the generation temperature to 0.0, frequency penalty to 0.0, and top-p to 1.0. Following the Text-to-SQL prompts proposed in~\cite{rajkumar2022evaluating, chen2023teaching}, we prepare various prompt templates for zero-shot and few-shot experiments by incorporating the DB schema including PK/FK key information.
The prompt templates used in experiments for both models can be found in our code repository.

\subsection{Impact of Design Dimensions}

We perform a detailed evaluation of the various Text-to-SQL systems according to the following dimensions (1) Data Model, (2) Language Model, (3) Training Data Size and (4) Pre- and Post-processing. The overall results for fine-tuned models and LLMs are summarized in Table~\ref{tab:model_accuracies_traditional} and Table~\ref{tab:model_accuracies_llms}, respectively.

\begin{table}[h]
\centering
\resizebox{1.0\linewidth}{!}{
\begin{tabular}
{l | c || r | r r }
\toprule
     \begin{tabular}[x]{@{}c@{}} \textbf{Data} \\ \textbf{Model} \end{tabular} & \begin{tabular}[x]{@{}c@{}} \textbf{Size} \\ \textbf{Train Set} \end{tabular}  & \textbf{ValueNet}  & \textbf{T5-Picard}
     & \textbf{T5-Picard$_{Keys}$}
     \\ 
     \toprule
     
     & zero & 2.00\%   & 8.00\%
     & 7.00\% \\
      & 100 &  16.00\%  & 22.00\%
      & 27.00\% \\
      \textbf{v1} & 200 & 18.00\%   & 29.00\%
      & 33.00\%\\
      & 300 &  20.00\%  &  29.00\%
      &  \textbf{38.00}\% \\
      \midrule
      & zero & 3.00\%  & 7.00\%
      & 7.00\%\\
      & 100 & 14.00\%  & 16.00\%
      & 29.00\%\\
      \textbf{v2} & 200 & 18.00\%  & 29.00\%
      & 33.00\%\\
      & 300  &  20.00\%  &  32.00\%
      &  \textbf{38.00}\% \\
      \midrule
      & zero  & 3.00\%  & 6.00\%
      & 8.00\%\\
      & 100  &  21.00\%  & 6.00\%
      & 25.00\%\\
      \textbf{v3} &  200 & 23.00\%  & 27.00\%
      & 36.00\%\\
      & 300 &  25.00\%    &  29.00\%
      &  \textbf{41.00\%}\\
\bottomrule
\end{tabular}
}

\caption{Execution accuracy of Text-to-SQL systems based on \emph{small to medium-size language models}. We apply train set sizes ranging from 100 to 300 samples for each system/data model combination.
}
\label{tab:model_accuracies_traditional}
\rmspace
\end{table}

\begin{table}[h]
\centering
\resizebox{1.0\linewidth}{!}{
\begin{tabular}
{l | c || r | c || r }
\toprule
     \textbf{Data Model} & \textbf{\#Shots} & \textbf{GPT-3.5}  & \textbf{\#Shots} & \textbf{LLaMA2-70B} \\ 
     \toprule
     
     & zero  & 25\% & zero & 5.00\%\\
      & 10  & \textbf{41.00}\%($\pm$3.40\%) & 2  & 11.25\%($\pm$2.77\%)\\
      \textbf{v1} & 20  & 39.00\%($\pm$4.08\%) & 4 & 10.50\%($\pm$3.35\%)\\
      & 30 &  37.00\%($\pm$2.49\%) & 8  & 16.00\%($\pm$3.67\%)\\
      \midrule
      & zero   & 25\% & zero  & 4.00\%\\
      & 10 & 37.00\%($\pm$2.83\%) & 2  & 8.75\%($\pm$2.68\%)\\
      \textbf{v2} & 20  & 36.00\%($\pm$1.63\%) & 4  & 8.50\%($\pm$1.50\%)\\
      & 30  & \textbf{37.50}\%($\pm$1.25\%) &  8  & 14.50\%($\pm$5.02\%) \\
      \midrule
      & zero   & 21\% & zero  & 5.00\%\\
      & 10 & \textbf{38.50}\%($\pm$5.31\%) & 2  & 8.50\%($\pm$1.80\%)\\
      \textbf{v3} &  20 & 37.00\%($\pm$2.49\%) & 4 & 8.50\%($\pm$3.20\%)\\
      & 30 & 37.00\%($\pm$1.89\%) &  8 &  15.00\%($\pm$4.47\%)\\
\bottomrule
\end{tabular}
}

\caption{Execution accuracy of Text-to-SQL systems based on \emph{large language models}.
For GPT-3.5, we used three different random samples of 10, 20, and 30 shots. Due to token size limits\tablefootnote{LLaMA2-70B has a limit of 4096 tokens.}, for LLaMA2 we performed experiments on up to 8 random samples with multiple folds.  For all systems, we report the mean accuracy and variance.
}
\label{tab:model_accuracies_llms}
\rmspace
\end{table}

\textbf{Data Model.}
How does the respective data model influence the execution accuracy of each Text-to-SQL system?
For the traditional fine-tuned systems in Table~\ref{tab:model_accuracies_traditional} we see that ValueNet actually improves its performance from v1 to v3 by \added{5\% points}  (for maximum training data size). However, T5-Picard exhibits little performance impact of the data model.
Upon investigation, we find that the reason for this behavior is that T5-Picard does not include PK/FK relationships and thus cannot take full advantage of the simplified data model and hence the simpler queries. For verification, we create a new T5 base model using a different encoding scheme from the one proposed in~\cite{raffel2020exploring}. The resulting T5-Picard$_{Keys}$ improves the performance greatly overall (up to 12\% points for v3), but more importantly shows now an increase of \added{3\% points} from v1 to v3, supporting our hypothesis. 

\added{Though ValueNet benefits from the model development, T5-Picard with keys still surpasses it and even ties with GPT-3.5 for data model V1. In our dataset v1 and v2, we have kept samples that cannot be answered by ValueNet. Thus, our experiment aims for fairness across all systems.
For the LLMs GPT-3.5 and LLaMA2-70B depicted in Table~\ref{tab:model_accuracies_llms}, we can observe no large differences between data models, with v1 showing the best overall results.
}

\textbf{Language Model.}
From the perspective of LLMs, we can clearly see that GPT-3.5, the largest language model with 175 billion parameters, outperforms, LlaMA2-70B, with 70 billion parameters, across the different data models. This outcome aligns with expectations as LlaMA2-70B has significantly fewer parameters and its token limitation constrained the number of few shot examples. 

T5-Picard$_{Keys}$ performs on par with GPT-3.5, both achieving 41\% accuracy. This is noteworthy because of the disparity in the number of parameters of each model, T5-Picard$_{Keys}$ only has 3 billion parameters. We also note the approximate 10\% gap in the performance of the T5-Picard model (which does not use the database schema) indicating that the schema information, and especially the foreign keys are a crucial input for the model enabling it to understand the database schema of a given query. 
Overall, given that the highest execution accuracy score of any system is only 41\%, the problem of translating natural language questions to SQL with large language models in a real-world application and real user queries is far from being solved. 

\textbf{Train Set Size.}
In terms of train set size, we observe that more training data typically leads to higher accuracy of the respective Text-to-SQL systems. For ValueNet, T5-Picard and T5-Picard$_{Keys}$, we used up to 300 training data samples.  
An important question in Text-to-SQL systems in practice is how much additional training data would actually improve their accuracy. Here, the trade-off is between manual annotation cost and the benefits of a better model.
To further investigate this trade-off, we train ValueNet with a total of 895 clean training samples \added{of the 1K labeled samples} for data model v3. \added{Note that we can only use 895 samples as the remaining 105 cannot be processed by the Spider SQL parser which is an essential part of the pre-processing component of ValueNet.}
For ValueNet, the accuracy increases from 25\% to about 29\%, i.e., tripling the amount of training samples shows an increase of \added{4\% points}. Hence, adding even more training data might not considerably increase these systems' accuracy.

For GPT-3.5, due to the token size limit, we provide a maximum of 30 random samples with each prompt. However, these 30 random samples are sufficient to perform as well as T5-Picard$_{Keys}$, which has a post-processing step, and has been fine-tuned on 300 samples. For a robust evaluation, we draw three different sample folds of up to 30 NL/SQL pairs for our few-shot experiments and report both the mean accuracy and the variance. Note the high variance, especially for the 10- and 20-shot experiments, indicates a high dependency on the specifically chosen samples. 

However, adding more training data also requires manual labeling effort and more computing resources.
We estimate that we used $\approx 1$ person month only for manual query annotation. 
The training compute time of ValueNet for all experiments took 182 hours on a single nVidia v100-32GB GPU. Fine-tuning for T5 on 4 v100-32GB GPUs took 88.4 hours, which is equivalent to 354 hours on a single GPU. 
While GPT-3.5 does not require training, running our few-shot experiments cost about 100 US dollars.
In summary, training with medium and large language models is both resource intensive but also has considerable monetary costs when performed on a large scale. Hence, deploying such systems in the real world often requires a trade-off between training data set size and the accuracy of the respective systems.

\textbf{Pre- and Post-processing.}
The complexity of pre-processing depends on the number of input parameters of each language model (see Table~\ref{tab:lm_dimensions}). For instance, ValueNet uses an intermediate representation, DB schema linking, and the database content as input. Hence, the pre-processing pipeline of ValueNet is more complex than the other Text-to-SQL systems, which only use conventional string normalization.
Due to ValueNet's utilization of an intermediate representation layer as output, a designated post-processing step is essential for converting the intermediate representation into a structured SQL query.
As we see from the performance of T5-Picard$_{Keys}$, adding FK information improves the performance of T5-Picard by 12\% in the v3 version of the data model indicating that the limitations of language models can be overcome by more informative input encodings.

Pre-processing for LLMs is less relevant,
because their comprehensive input prompts can include FK information and sample rows 
(as we demonstrate in our experimental evaluation).
We leave the exploration of post-processing strategies LLMs to future work.

\subsection{Impact of Query Complexity and Query Characteristics}
We now investigate the impact of (1) the SQL query complexity and (2) certain query characteristics on the execution accuracy. For these experiments, we only use the models trained with the maximum training data size, i.e. 300 train set samples for ValueNet, T5-Picard and T5-Picard$_{Keys}$ as well as 30 shots for GPT-3.5 and 8 shots for LLaMA2-70B.

\begin{figure}[t]
    \centering
    \includegraphics[width=1.0\columnwidth]{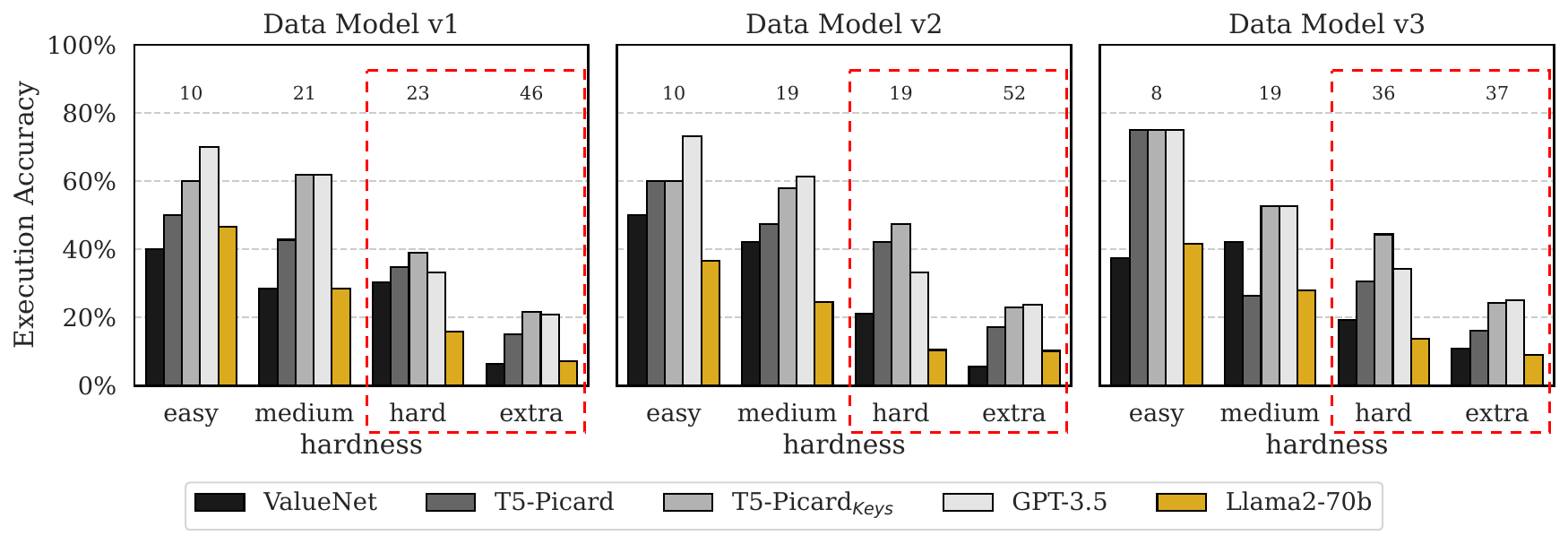}
    \caption{Execution accuracy of Text-to-SQL systems per Spider hardness level. The numbers on top of the bars represent the number of test set queries in each category.}
    \label{fig:data_model-system-hardness}
    \rmspace
\end{figure}

\textbf{Query Complexity.} First, we find that increasing Spider hardness reduces execution accuracy regardless of the Text-to-SQL system and data model. E.g., the accuracy for easy queries reaches up to 77\%, while for extra hard queries the highest accuracy is slightly above 20\% across all data models and systems (see Figure~\ref{fig:data_model-system-hardness}).
We can directly \emph{see the impact of the data model on the number of extra hard queries}, which change from 46 (v1) over 52 (v2) to only 36 in v3 (see red dashed lines). We find that the \emph{hardness of these queries reduces from extra hard to hard} (36 in v3) with no impact on the execution accuracy for hard queries.

\begin{figure}[ht]
    \centering
    \includegraphics[width=1.0\columnwidth]{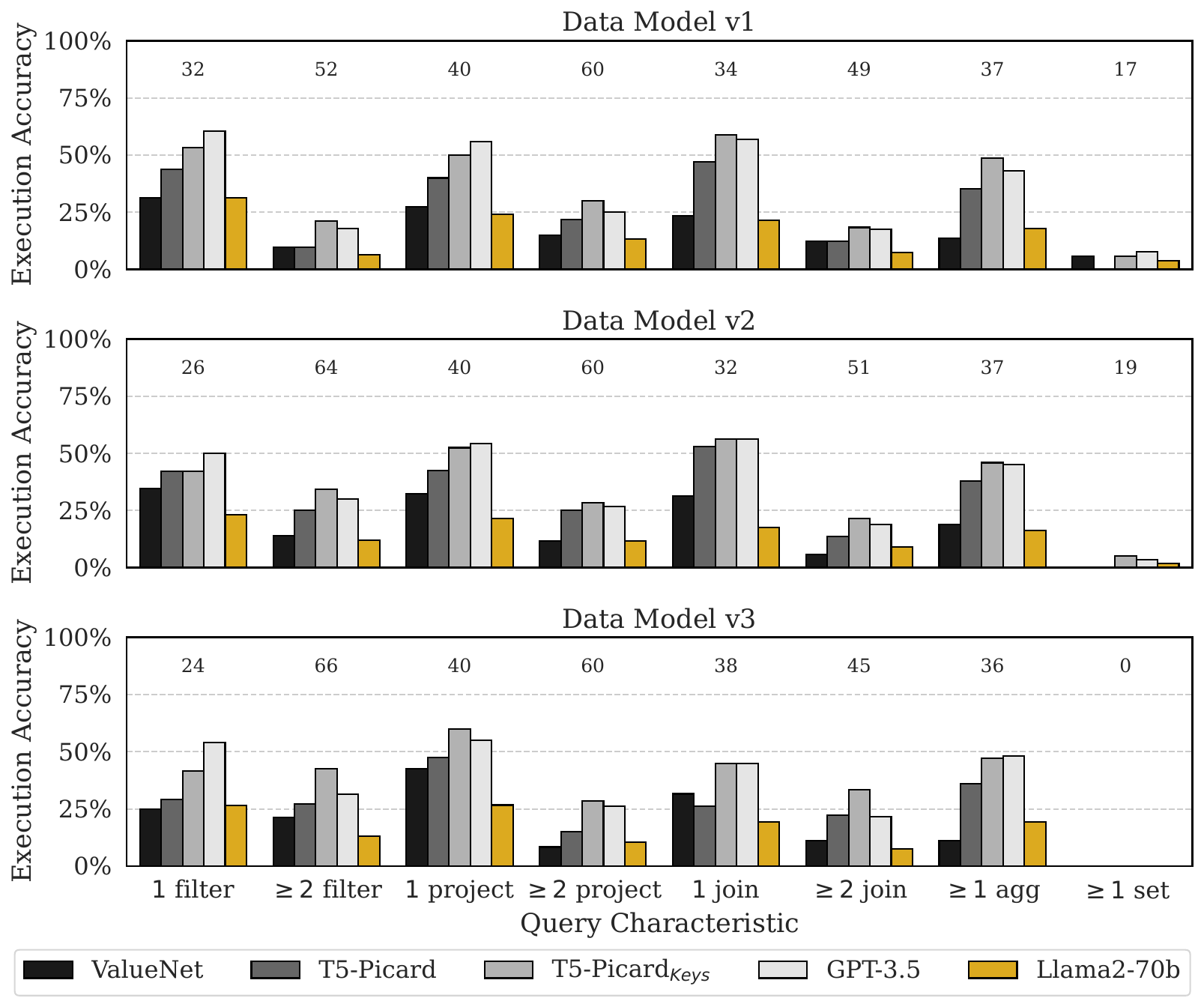}
    \caption{Execution accuracy per query characteristic. The numbers on top of the bars represent the number of test set queries in each category.
    }
    \label{fig:sql-query-characteristic-without-subquery-300-data-model-v1-v2-v3}
    \rmspace
\end{figure}

\textbf{Query Characteristics.} To further investigate the impact of query complexity, we also evaluate the accuracy of our defined query characteristics from Table~\ref{tab:train_dev_data_statistics_transposed}, i.e. the number of \emph{joins}, \emph{projections filters}, etc. (see Figure~\ref{fig:sql-query-characteristic-without-subquery-300-data-model-v1-v2-v3}). Except for projections and aggregations, we observe some interesting variances in accuracy between the data models for certain query characteristics:

\emph{Filters.} Here the number of queries with only one filter reduces from 32 in v1 to 24 in v3, while the accuracy for these queries reduces slightly across all Text-to-SQL systems. Contrary to that, the number of queries with $\geq 2$ filters increases from 52 to 66, while the accuracy of these queries also increases across all systems.  \added{This indicates that filters are intuitive in Text-to-SQL, i.e. for a column like ``winner'' the model expects a Boolean value,  (see Listing~1).}
\begin{lstlisting}[language=SQL,basicstyle=\footnotesize\ttfamily\color{black},breaklines=true, caption={\added{Query for ``How many times did England win the worldcup?'' (top for v1, bottom for v3)}},captionpos=b]
SELECT count(*)
FROM world_cup AS T1
JOIN national_team AS T2 ON T1.winner = T2.team_id
WHERE T2.teamname = 'England';
SELECT count(*)
FROM world_cup_result AS T1
JOIN national_team AS T2 ON T1.team_id = T2.team_id
WHERE T2.teamname = 'England' and T1.winner = 'True';
\end{lstlisting}



\emph{Joins.} We see most $\geq 2$ joins for data model v2, while v3 requires the least. Investigating the affected queries, we observe that they only require a single join in v1, which is why the number of single join queries increases from v2 to v3. Unfortunately, these single-join queries are still hard for most Text-to-SQL systems. Only ValueNet and Llama2 can sustain their accuracy from v2 to v3.

\emph{Sets.} The maybe most impactful effect of our data model changes can be observed from the number of set queries.
Set operations, which showed poor performance across all evaluated systems, are no longer needed in the v3 data model, improving overall accuracy.

Overall, we observe that different data models can change query characteristics in various ways. Towards an explicit Text-to-SQL data model design strategy, \emph{emphasis should be especially put on characteristics that are only poorly supported, such as set operations}. If these can be avoided in a different data model design, the overall performance of Text-to-SQL system can improve without additional training data or more advanced language models.

\subsection{Inference Time of Language Models}

Table \ref{tab:inference_speed} shows the mean and standard deviation of the inference time of each Text-to-SQL system in seconds per query, including the hardware specifications for each of the fine-tuned models and the open source, LlaMA2-70B. For ValueNet and all of the T5 models we used Tesla v100-SXM2-32GB GPUs. LlaMA2-70B requires at least 35GB of GPU memory with 8bit quantizing\cite{dettmers2022optimizers}, necessitating the use of the more powerful Tesla A100-SXM4-40GB GPU.  

We see that for ValueNet, which is based on a small language model with 148M parameters, the mean inference time per query is about 1 sec. We note that for T5-Picard and T5-Picard$_{Keys}$, which have 3B parameters, the mean inference time per query is 652 and 294 seconds per query, respectively. In other words, \emph{it takes more than 10 minutes (!) for T5-Picard to translate an NL question to a SQL query}. Such a response time is impractical for usage in a real-world scenario without considerable hardware ramp-up or substantial performance optimizations.

For GPT-3.5 the mean inference time per query is 2.51 seconds. However, note that GPT-3.5 is run on a massive hardware infrastructure in the cloud under the control of OpenAI. 

Finally, the mean response time per query for LLaMA2-70B is 37.03 seconds. It is noteworthy to consider that this model requires a significantly more powerful hardware setup of 4 A100 GPUs compared to the setup of ValueNet and T5-Picard with only 1 v100 GPU. In summary, LLaMA2-70B's inference time of more than half a minute per query to translate Text-to-SQL cannot be used practically without a considerable hardware investment.

\begin{table}[hb]
\centering
\resizebox{1.0\linewidth}{!}{
\begin{tabular}
{l || r | r r | r | r }
\toprule
    & \textbf{ValueNet} & \textbf{T5-Picard}  
    & \textbf{T5-Picard$_{Keys}$} 
     & \textbf{GPT-3.5} & \textbf{LLaMA2-70B} \\ 
     \toprule
     
     \textbf{Time (sec)}&1.06 $\pm$0.14  & \textbf{652.16} $\pm$165.94
     & 294.08 $\pm$75.65 & 2.51 $\pm1.06$& 37.03 $\pm17.30$\\ 
    \textbf{Hardware}&v100
    &v100 &v100 & - & A100 \\ 
    \textbf{\# GPUs}& 1 &1
    & 1& -&4 \\ 
\bottomrule
\end{tabular}
}

\caption{Mean and standard deviation of the inference unit time of each Text-to-SQL system in seconds per query.}
\label{tab:inference_speed}
\rmspace
\end{table}
\section{Discussion and Lessons Learned}
\label{sec:discussion}

From our in-depth design space evaluation, we learned several important lessons that can guide  designs of Text-to-SQL systems.

\noindent \textbf{Data Models Matter -- Sometimes}
For the small- and medium-sized models, our results underscore the \emph{significance of the keys' information, i.e., using the primary keys and foreign keys as input to the language model} to take advantage of the characteristics of the underlying data model. The results from T5-Picard and T5-Picard$_{Keys}$ show that including keys' information substantially enhances the execution accuracy across all data models in all experiments with various training data sizes. This enhancement is more significant on optimized data models with smaller training data sizes, yielding an improvement of up to 19\% (see Table \ref{tab:model_accuracies_traditional}). 

Although ValueNet and T5-Picard$_{Keys}$ distinguish each other in many specifications, e.g., model architectures or model sizes, they share a common requirement for keys' information as input and a common step of post-processing before prediction. Hence, both models clearly manifest the benefits from data model optimization with each training data size. 
\added{In summary our results show that \emph{LLMs are robust to the changes and design space parameters} we evaluated while small and medium sized language models are not.
}

\noindent \textbf{Model Inference Time Matters.}
Even though T5-Picard$_{Keys}$, which uses a medium-size language model of 3B parameters, has a comparable execution accuracy to GPT-3.5, it's \emph{practical usage is currently limited given an inference time of more than 5 minutes per query}. In order to use that system in practice, a considerable hardware investment is required.

\added{
\noindent \textbf{Evaluating Text-to-SQL in Practice.}
We discover several challenges wrt evaluating Text-to-SQL systems with real user queries in a practical application. Some of these, such as the ambiguity of natural language or the semantic mismatch between user request and database content, have also been recently discussed in~\cite{floratou2024nl2sql}.
This also related to the question of developing a good evaluation methodology, with important decisions such as train/test split~\cite{finegan2018improving}.
Here, our diversity-based sampling strategy results in a challenging dataset, where even the best method achieves only 41\%).
}

\added{
\noindent \textbf{Limitations.}
A key reason for the changes in the schemas in v1, v2, and v3 was to facilitate query writing. As seen in Figure \ref{fig:sql_sample}, the SQL queries for v1 and v2 are very long and difficult to write manually, while the v3 queries are much more intuitive and easier to write.}
\addedd{While we adapted the schema design based on the questions submitted, the impact of changing the database schema could be further explored from the natural language point of view by employing lexical variation, leading to better alignment between the natural language questions and the database schema. 
}

\addedd{
A possible limitation of the evaluation's validity comes from the lack of full independence between the choice of the natural language queries used in our evaluation and the three data models. First, some of the queries submitted to FootballDB were themselves used to adjust the schema in producing data models v2 and v3.
Second, we used the Spider hardness of queries in data model v3 to sample our train and test sets for data models v1 and v2, influencing the 
distribution of queries by the query hardness in v3.
Finally, applying experiments to a single database for all insights into data model refactoring might be insufficient to demonstrate the generalizability of the impact of data model optimization across diverse databases. Additional experiments on more databases are necessary to substantiate the applicability of these optimizations and draw more general conclusions. 
}
\section{Related Work}
\label{sec:related_work}

Throughout the paper we have mentioned various Text-to-SQL systems. For a general overview we refer the reader to~\cite{gkini2021depth, kim2020natural, affolter2019comparative}.
To the best of our knowledge there has not been any practical exploration of Text-to-SQL systems such as we presented in this paper.
However, there have been several community driven datasets and benchmarks that have been crucial to the advancement in Text-to-SQL systems that can be classified into cross-domain and domain-specific:

WikiSQL~\cite{zhongSeq2SQL2017} and Spider~\cite{yu2018spider} have been two of the most widely used cross-domain benchmark datasets and helped to show the potential of deep learning for Text-to-SQL.
ScienceBenchmark~\cite{zhang2023sciencebenchmark} contains databases from astrophysics, bioinformatics and policy research. The main challenge of this benchmark is the complexity of the domains which require expert knowledge.
BIRD~\cite{li2023can} includes 95 databases across 37 domains created via crowdsourcing. BIRD also introduces a new evaluation metric called Valid Efficiency Score.

Also, domain-specific datasets have been created in various domains, encompassing topics such as entertainment~\cite{sqlizer, deriu2020methodology}, consumer reviews~\cite{sqlizer}, medical data~\cite{wang2020medical}, and community-generated Q\&A data~\cite{hazoom2021wildtext-to-sql}.
These datasets provide a valuable play-ground for building domain specific Text-to-SQL systems.

Compared to above works, the unique value of FootballDB is that its data is based on a \emph{months-long live deployment with $\approx$6K real user questions}. Moreover, we developed three data models for our dataset, for which we manually annotated 1,200 NL/SQL pairs.
The effect of various data models and different sizes of training data on the accuracy of Text-to-SQL systems has not been studied before to our knowledge.
\added{
Table~\ref{tab:data_schema_compare} summarizes the main relevant existing Text-to-SQL benchmark datasets. FootballDB is the only dataset that allows a multi-schema evaluation, it is the only dataset based on real-user questions posed to a live system, and exhibits the most tokens per query (an indication of query complexity).
}

\begin{table}[h]
\setlength\tabcolsep{5pt}
    \centering   
    \added{
    \resizebox{\linewidth}{!}{
    \begin{tabular}{l|ccccc}
        \toprule
        \multirow{1}{*}{\textbf{Dataset}}  & \multirow{1}{*}{\textbf{\#Examples (\#DBs)}} & \multirow{1}{*}{\textbf{\#Tables (\#Rows)/DB}} & \multirow{1}{*}{\textbf{\#Tokens/Query}} & \multirow{1}{*}{\textbf{Multi-Schema}} & \multirow{1}{*}{\textbf{Live Users}} \\
        \midrule
        WikiSQL\cite{zhong2017seq2sql} & 80,654 (26,521) & 1 (17)  & 12.2  & 
        
        \textcolor{red}{\xmark} & \textcolor{red}{\xmark} \\
        SPIDER \cite{yu2018spider} & 10,181 (200) & 5.1 (2K)  & 
        18.5  & 
        \textcolor{red}{\xmark} & \textcolor{red}{\xmark} \\
        KaggleDBQA \cite{lee2021kaggledbqa} & 272 (8) & 2.3 (280K)  
        &
        13.8  & 
        \textcolor{red}{\xmark} & \textcolor{red}{\xmark} \\
        ScienceBench. \cite{zhang2023sciencebenchmark} & 5,332 (3)& 16.7 (51M)  &
        15.6  & 
        \textcolor{red}{\xmark} & (\textcolor{teal}{\cmark}) \\ 
        BIRD \cite{li2023llm} & 12,751 (95) & 7.3 (549K)  & 
        30.9  & 
        \textcolor{red}{\xmark} & \textcolor{red}{\xmark} \\ 
        \midrule
        \textbf{FootballDB} & \textbf{1200 (3)} & \textbf{15 (107K)}  &
        \textbf{33.7}  & 
        \textcolor{teal}{\cmark} & \textcolor{teal}{\cmark} \\
        \bottomrule
    \end{tabular}
    }}
     \caption{\added{Comparison between FootballDB and  existing Text-to-SQL datasets. 
     \texttt{Multi-Schema:} multiple schemas, same data.
     \texttt{Live Users:} real user questions against live systems.}}
     \label{tab:data_schema_compare}
     \rmspace
\end{table}
\section{Conclusion}

This paper provides essential insight into implementing Text-to-SQL systems in the real world including an in depth analysis of the impact of different data models.

First, we report on \emph{over nine months of deployment with several hundred users}, the issues we faced, and the design iterations implemented to address them, providing a unique perspective for Text-to-SQL practitioners from industry and academia.

Second, we take our practical deployment as an opportunity to define and \emph{evaluate critical design space dimensions of Text-to-SQL systems}, namely (1) Data Model, (2) Language Model, (3) Training Data Size and (4) Pre-/Post-processing. We pay particular attention to the data model impact which is a previously understudied subject.
Our experimental results show that the choice of data model has a substantial impact on the accuracy of Text-to-SQL systems when using small or medium-sized language models. Our experiments also demonstrate that medium-sized and large language models show the highest execution accuracy (T5-Picard$_{Keys}$ and GPT-3.5: 41\%) with little labeled data (10 few shot examples for GPT-3.5). 

Third, an important aspect that has not been given much attention yet is the \emph{inference time of Text-to-SQL systems}. Our experiments show that the mean inference time for T5-Picard$_{Keys}$ and LLaMA2-70B deployed on a reasonable GPU hardware is 294 and 37 seconds per query, respectively. These high response times make it currently impractical to use these systems in real-world scenarios where interactive query speeds of less than 3 seconds are expected. Reducing the maximum input token size has the potential to meet the inference time requirements. Nevertheless, this reduction can be accompanied by a significant decrease in the execution accuracy of the predicted SQL query due to the lossy input information. Hence, considerable hardware investment or algorithmic optimizations would be required to make these systems practical.

Last, we \emph{release FootballDB, a new benchmark dataset} based on over 6K user queries with 1200 manual annotated NL-SQL pairs for three distinct data models.
We aim to extend FootballDB with a hidden test dataset and release a public benchmark in the same vein as the Spider~\cite{yu2018spider} and  BIRD~\cite{li2023can} benchmarks
to support other researchers in developing and testing their systems on an internationally well-known and easily understandable domain.

\begin{acks}
This project has received funding from the European Union’s Horizon 2020 research and innovation program under grant agreement No 863410. We also want to thank Jan Deriu and Katsiaryna Mlynchyk for valuable contributions to the FootballDB.
\end{acks}

\balance

\bibliographystyle{ACM-Reference-Format}
\bibliography{bibliography}
\end{document}